\documentclass[a4paper]{jpconf}
\usepackage{graphicx, color}
\begin{document}
\title{Electromagnetic emission from hot medium measured by the PHENIX experiment at RHIC}

\author{Takao Sakaguchi, for the PHENIX collaboration}

\address{Brookhaven National Laboratory, Physics Department, Upton, NY 11973, USA}

\ead{takao@bnl.gov}

\begin{abstract}
Electromagnetic radiation has been of interest in heavy ion collisions because
they shed light on early stages of the collisions where hadronic probes do
not provide direct information since hadronization and hadronic interactions
occur later. The latest results on photon measurement from the PHENIX
experiment at RHIC reflect thermodynamic properties of the matter produced
in the heavy ion collisions. An unexpectedly large positive elliptic flow
measured for direct photons are hard to be explained by many models.
\end{abstract}

\section{Introduction}
\begin{figure}[h]
\begin{minipage}{18pc}
\includegraphics[width=17.5pc]{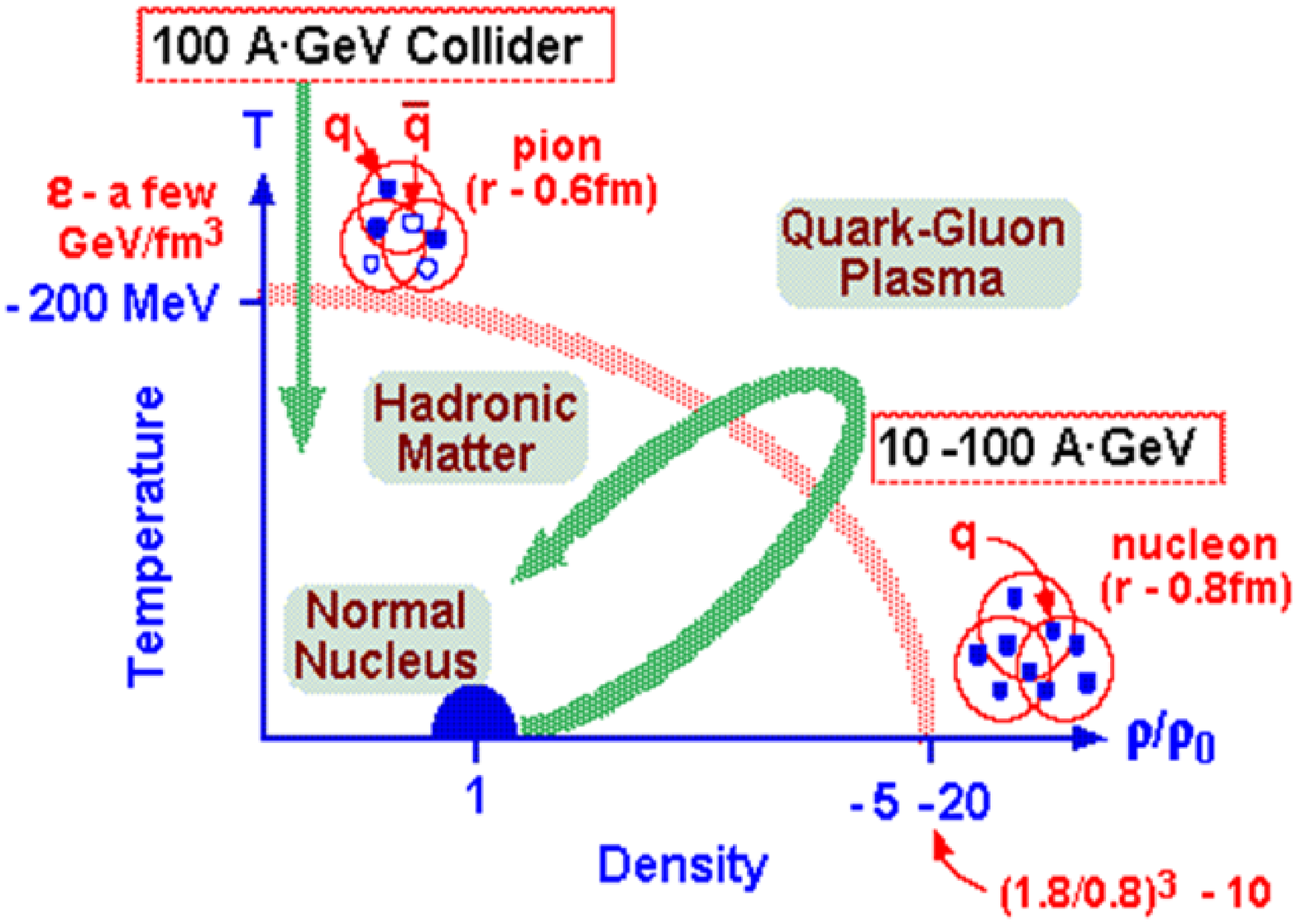}
\caption{\label{fig1}Phase diagram of the nuclear matter.}
\end{minipage} \hspace{2pc}%
\begin{minipage}{18pc}
\includegraphics[width=17.5pc]{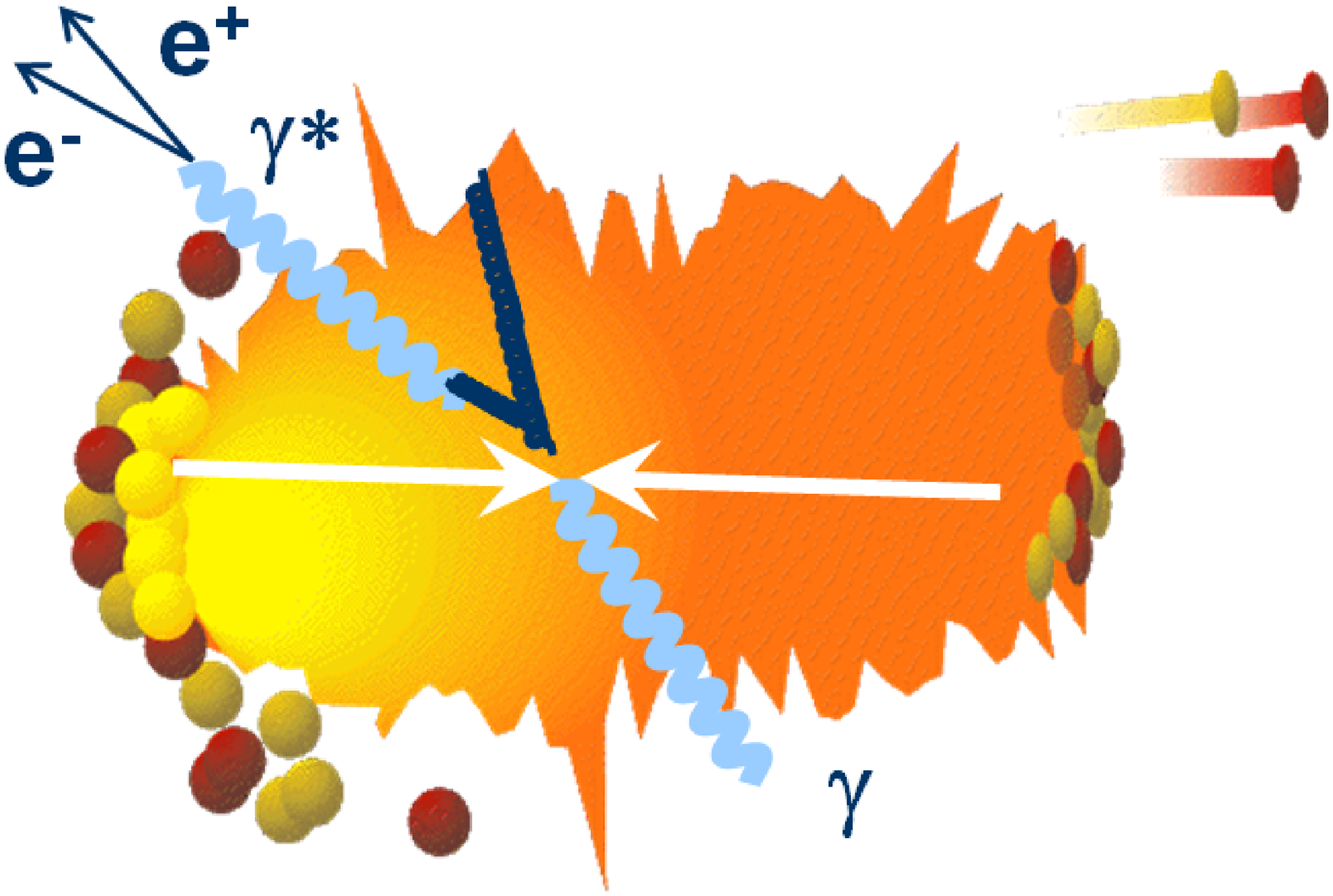}
\caption{\label{fig2}Photon emission in relativistic heavy ion collisions.}
\end{minipage}
\end{figure}

The experiments utilizing relativistic heavy ion collisions have been
aiming to find a new state of matter, quark-gluon plasma (QGP),
that should have existed in the early stage of the Universe (Fig.~\ref{fig1}).
The QGP is an interesting state in the sense that it is not only a
discovery subject, but also a unique place to understand the nature
of QCD matter, such as quark confinement or the chiral symmetry
restoration. The unique feature of the study at the Relativistic Heavy
Ion Collider (RHIC) at the Brookhaven National Laboratory is that one can
utilize the probe with high $Q^2$ (perturbative probe) to investigate
the QCD matter in thermal region (low $Q^2$, non perturbative matter).

Many intriguing phenomena have been observed at RHIC since it started
of running in 2000. The high transverse momentum ($p_T$) hadron production
from the initial hard scattering was observed, and the large suppression
of their yields suggested that the matter is sufficiently dense to 
cause parton-energy loss prior to hadronization~\cite{Adcox:2004mh}.
The large elliptic flow of particles and its scaling in terms of particle
kinetic energy suggests that the system is locally in equilibrium as early as 
0.3\,fm/c, and the flow occurs already on the partonic level.

\begin{figure}[h]
\begin{minipage}{18pc}
\includegraphics[width=17.5pc]{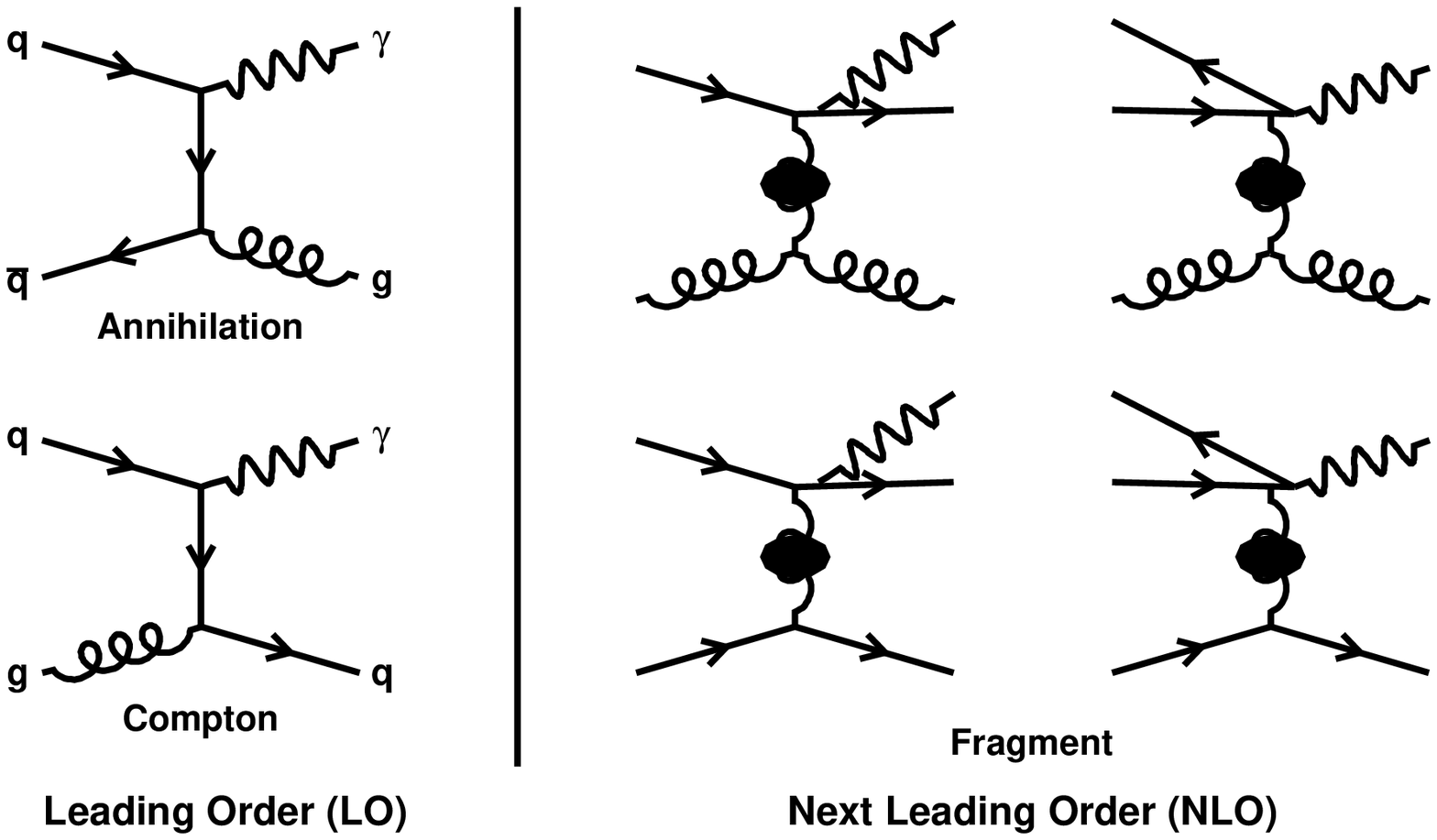}
\caption{\label{fig3}Photon production process.}
\end{minipage} \hspace{2pc}%
\begin{minipage}{18pc}
\includegraphics[width=17.5pc]{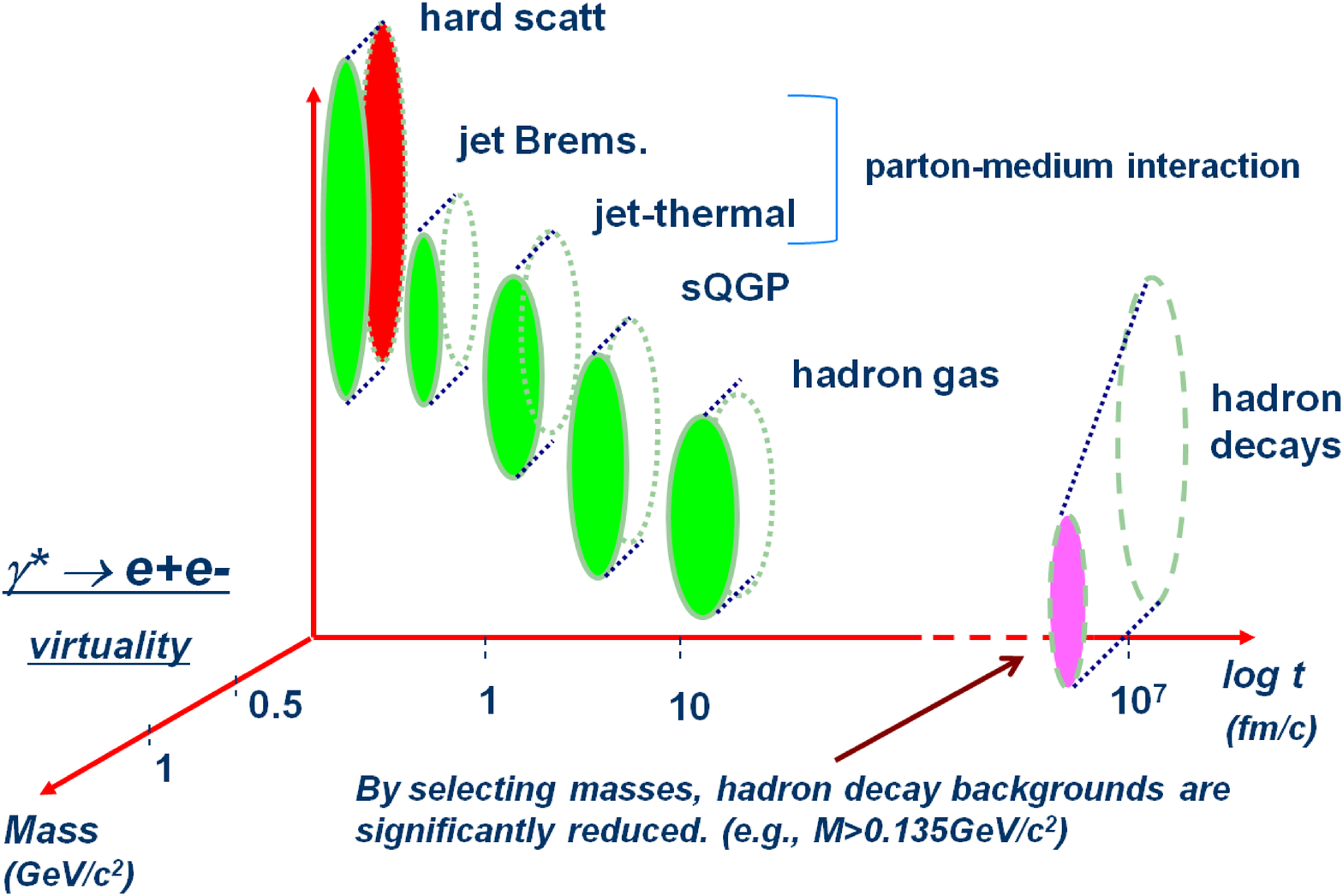}
\caption{\label{fig4}Sources of photons from various stages of collisions.}
\end{minipage}
\end{figure}

Because they interact with the medium and other particles only
electromagnetically and are largely unaffected by final state interactions,
photons serve as a direct and penetrating probe of the early stages at high
temperature and high density~\cite{Stankus:2005eq}.
At leading order, the production processes of photons are annihilation
($q\bar{q}\rightarrow\gamma g$) and Compton scattering 
($qg \rightarrow \gamma q$) (Figure~\ref{fig3}). Their yields are
proportional to $\alpha\alpha_{s}$, which are $\sim$40 times lower than
hadrons from strong interactions.

A calculation predicts that the photon contribution from the QGP state is
predominant in the $p_T$ range of 1$<p_T<$3\,GeV/$c$~\cite{Turbide:2003si}.
For $p_T>$3\,GeV/$c$, the signal is
dominated by a contribution from initial hard scattering, and $p_T<$1\,GeV,
the signal is from hadron gas through processes of
$\pi\pi(\rho) \rightarrow \gamma \rho(\pi)$, 
$\pi K^* \rightarrow  K \gamma$ and etc. Figure~\ref{fig4} shows a landscape
of photon sources as a function of the time they are produced. The vertical
axis corresponds to transverse momenta of photons. We have one another degree
of freedom, virtual mass, in photon measurement, which will be explained in
detail in a later section.
These photons can be measured after a huge amount of background photons
coming from hadron decays ($\pi^0$, $\eta$, $\eta'$ and $\omega$, etc.)
are subtracted off from inclusive photon distributions. The typical
signal to background ratio is $\sim$1\,\% at 2\,GeV, and $\sim$10\,\% at
5\,GeV in case of p+p collisions. The signal from QGP is predicted to be
$\sim$10\,\% of the inclusive photons.
For Au+Au collisions, thanks to a large suppression of high $p_T$ hadrons,
the ratio is enhanced by the same degree.
PHENIX~\cite{Adcox:2003zm} has measured photons throughout the first decade
of RHIC operations. We present here a review of the results.

\section{Measurement of initial hard scattering photons in heavy ion collisions}
\begin{figure}[h]
\begin{minipage}{18pc}
\includegraphics[width=17.5pc]{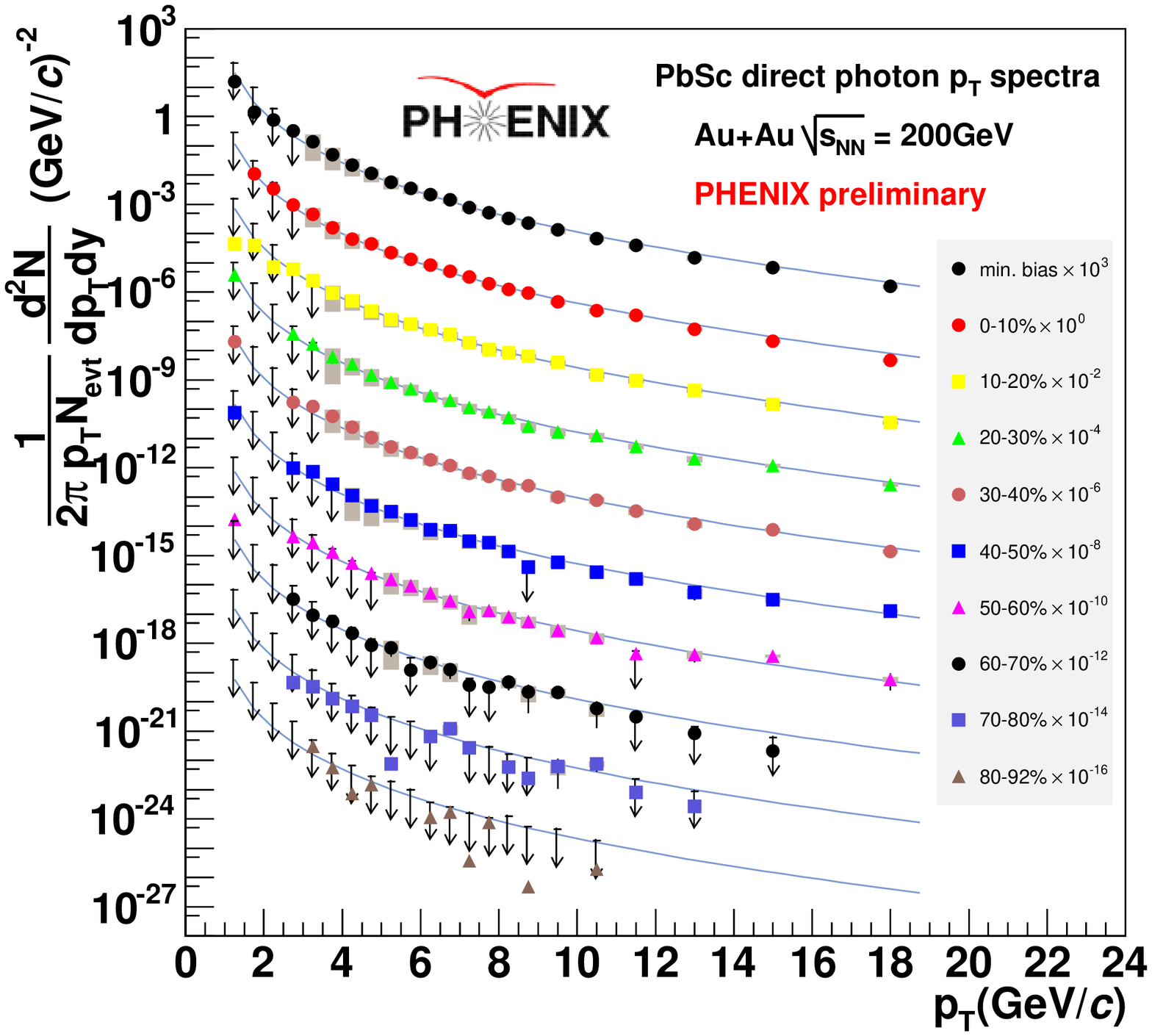}
\caption{\label{fig5}Direct photon spectra in Au+Au collisions at $\sqrt{s_{NN}}$=200\,GeV.}
\end{minipage} \hspace{2pc}%
\begin{minipage}{18pc}
\includegraphics[width=17.5pc]{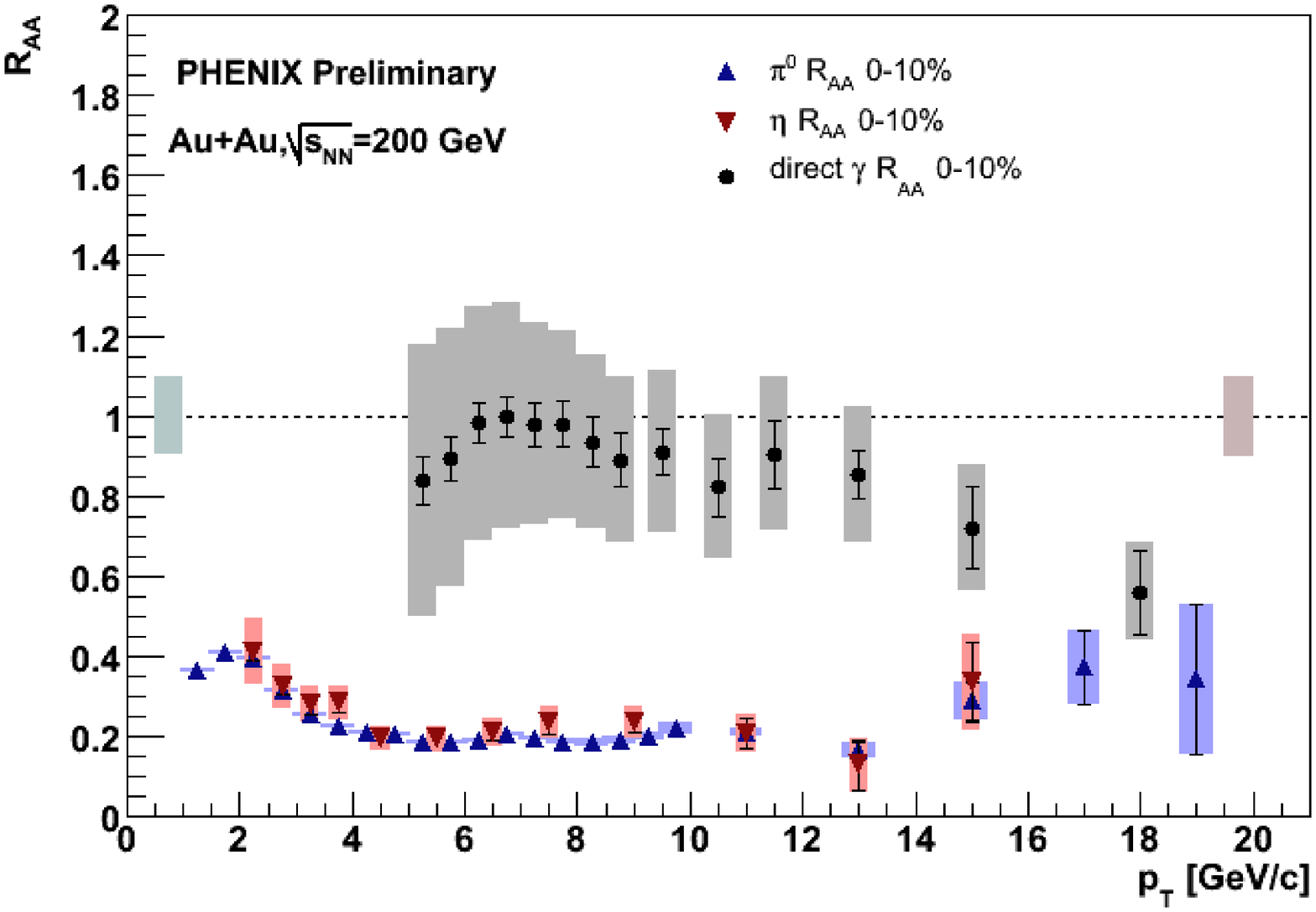}
\caption{\label{fig6}Nuclear modification factors ($R_{AA}$ for photons, $\pi^0$ and $\eta$ in 10\,\% central Au+Au collisions at $\sqrt{s_{NN}}$=200\,GeV.}
\end{minipage}
\end{figure}

One of the big successes by now in electro-magnetic radiation measurements is
the observation of high $p_T$ direct photons that are produced in initial
hard scattering~\cite{Adler:2005ig} in relativistic heavy ion collisions.
Figure~\ref{fig5} shows the direct photon spectra in Au+Au collisions
at $\sqrt{s_{NN}}$=200\,GeV for different centralities.
The lines show the NLO pQCD calculations~\cite{Gordon:1993qc} scaled by
the nuclear thickness
function ($T_{AA}$). The fact that the data are well described by the lines
show that the yields are following the $T_{AA}$ scaling and suggest that
the source is the initial hard scattering.
Figure~\ref{fig6} shows the nuclear modification factors ($R_{AA}$) for direct
photons, $\pi^0$ and $\eta$ for 0-10\,\% central Au+Au collisions at the same
center-of-mass (cms) energy. $R_{AA}$ is defined as the ratio of the yield in
nucleus-nucleus collisions divided by that in p+p collisions scaled by $T_{AA}$.
The high $p_T$ hadron suppression is interpreted as a consequence of an
energy loss of hard-scattered partons in the hot and dense medium.
It was strongly supported by the fact that the high $p_T$ direct photons are
not suppressed and well described by a NLO pQCD calculation. The small
suppression seen in the highest $p_T$ is likely due to the fact that the ratio
of the yields in Au+Au to p+p was computed without taking the isospin
dependence of direct photon yields into account~\cite{Arleo:2006xb}.

\section{Measurement of direct photons through its internal conversion}
\begin{figure}[h]
\begin{center}
\includegraphics[width=32pc]{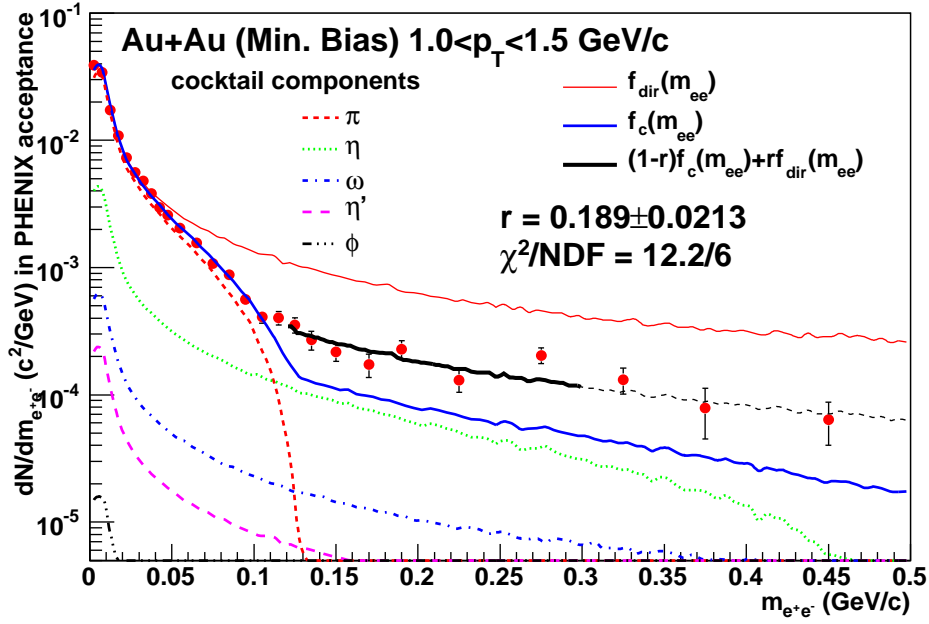}
\caption{\label{fig7}Invariant mass distributions of electron-pairs and comparison with possible hadron sources of electron-pairs.}
\end{center}
\end{figure}

\begin{figure}[h]
\begin{minipage}{18pc}
\includegraphics[width=17.5pc]{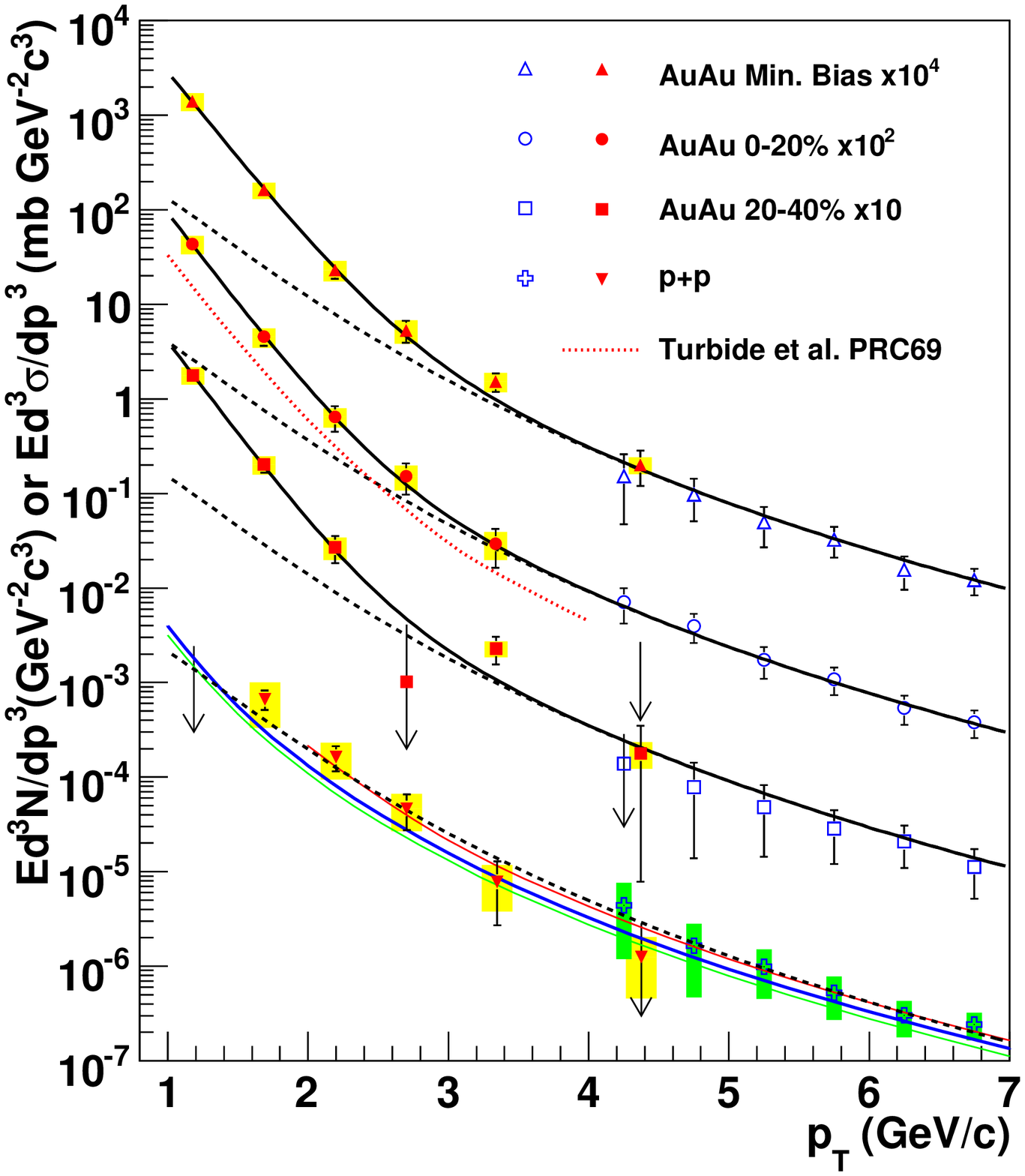}
\caption{\label{fig8}Direct photon spectra obtained from the measurement of internal conversion of photons in Au+Au collisions.}
\end{minipage} \hspace{2pc}%
\begin{minipage}{18pc}
\includegraphics[width=17.5pc]{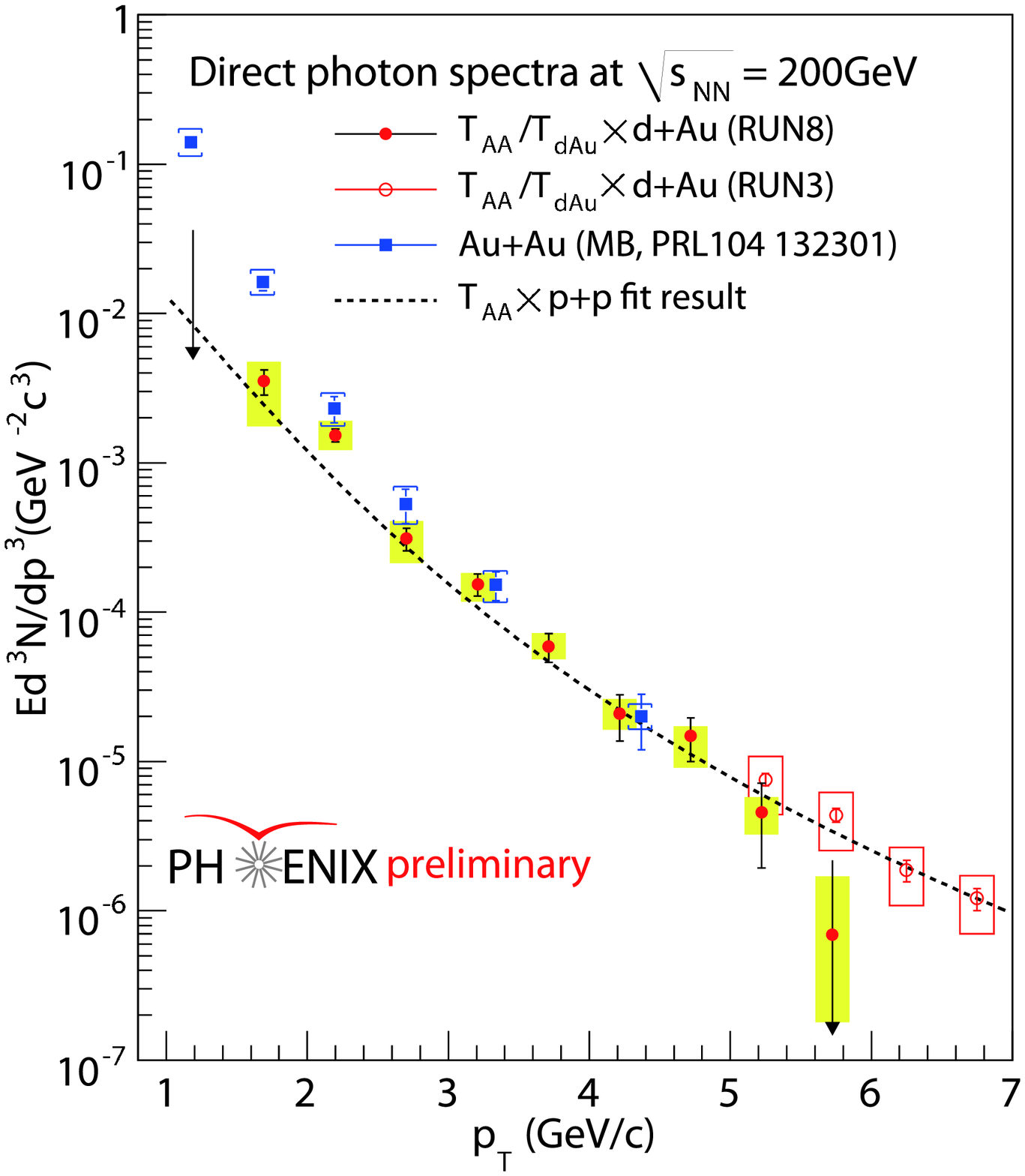}
\caption{\label{fig9}Direct photon yield in Au$+$Au and $d$$+$Au collisions scaled by the difference of $N_{\rm coll}$.}
\end{minipage}
\label{fig_dAuSpec2}
\end{figure}

There is a huge background arising from $\pi^0$ decaying into two
photons, which makes it very difficult to look at the direct photon signal
at low $p_T$, where thermal photons from QGP manifest, with traditional
calorimetry of (real) photons. However, if we look at photons with a small
mass (virtual photons) instead, we can select the mass region where $\pi^0$
contribution ceases (Fig~\ref{fig7}). For the case of $p_T>>M$, the yield of
virtual photons is expected to be dominated by internal conversion of real
photons~\cite{Adare:2008fqa,Adare:2009qk}. 
For obtaining direct photon yield, we fit the measured invariant mass
distribution with the function:

\[ F = (1-r) f_c + r f_d,\]

\noindent where $f_c$ is the cocktail calculation (photons from various
hadron decays), $f_d$ is the mass distribution for direct photons,
and $r$ is the free parameter in the fit. 
Next, using the Kroll-Wada formula~\cite{Kroll:1955zu} to account 
for the Dalitz decays of $\pi^0$, $\eta$ and direct photons, $r$ is 
defined as the ratio of direct photons to inclusive photons:
\[ r = \frac{\gamma^{*}_{\rm dir} 
(m_{ee}>0.15)}{\gamma^{*}_{\rm inc}(m_{ee}>0.15)} \propto \
 \frac{\gamma^{*}_{\rm dir} (m_{ee}\approx 0)}
{\gamma^{*}_{\rm inc}(m_{ee}\approx 0)} \
= \frac{\gamma_{\rm dir}}{\gamma_{\rm inc}} \equiv r_{\gamma} \]
Then, the invariant yield of direct photons is calculated as 
$\gamma_{\rm inc} \times r_{\gamma}$.  
As described in~\cite{Adare:2009qk}, the procedure is demonstrated in 
Fig~\ref{fig7} for 1.0$<p_T<$1.5\,GeV/$c$.
The dotted lines show the contributions from various hadrons, the solid blue
is the sum of these contributions, and the solid red line shows the
distribution from direct photons converted internally.  The $r$ value is
determined by the fit to the data. The error of the fit corresponds to the
statistical error. We applied the procedure as a function of $p_T$ for various
centrality selections in p+p and Au+Au collisions, and obtained the $p_T$ 
spectra, as shown in Fig~\ref{fig8}.
The distributions are for 0--20\,\%, 20--40\,\% centrality and MB events 
for Au$+$Au collisions. For $p_T<$2.5\,GeV/$c$ the Au+Au yield are visibly
higher than the scaled p+p yield. The distributions were then fitted with
the p+p fit plus exponential function to obtain slopes and dN/dy for three
centralities. The slopes are estimated to be $\sim$220\,MeV. The lines show
the theoretical expectation from a literature~\cite{Turbide:2003si}.
One may question whether or not the excess arises from a source that 
exists only in Au$+$Au collisions. For example, an effect that could 
increase the yield is cold-nuclear-matter (CNM) effect such as $k_T$ 
broadening (Cronin effect).  To quantify the contribution we analyzed
2008 $d$$+$Au data with the same procedure~\cite{Sakaguchi:2010hx}.
Figure~\ref{fig9} shows the Au$+$Au yield compared to the $d$$+$Au
yield scaled by $N_{\rm coll}$. It clearly shows that there is an
enhancement over CNM effects in Au+Au collisions.

\section{Exploring new degree of freedom in direct photon measurement}
On exploring the matter produced, one wants to explore a new degree of
freedom of the observables. The angular dependence of the photon yield
with respect to the plane defined by impact parameter (event plane) is
one of the degrees that can be investigated. Rapidity dependence will be
another degree of freedom, which may shed light to the pre-equilibrium
state of the collisions.
\begin{figure}[h]
\begin{minipage}{18pc}
\includegraphics[width=17.5pc]{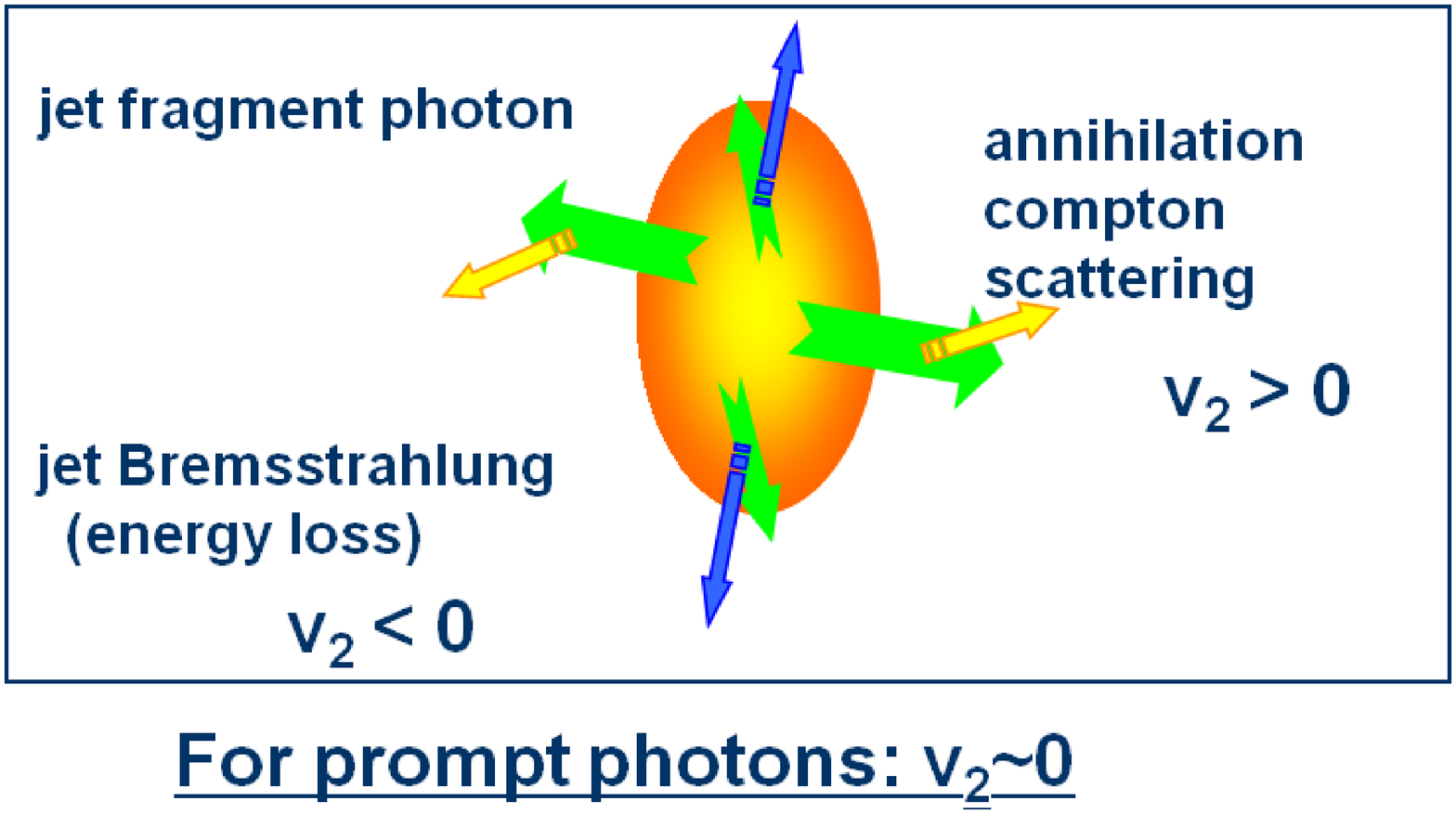}
\caption{\label{fig10} Source dependence of elliptic flow ($v_2$) of direct photons.}
\end{minipage} \hspace{2pc}%
\begin{minipage}{18pc}
\includegraphics[width=17.5pc]{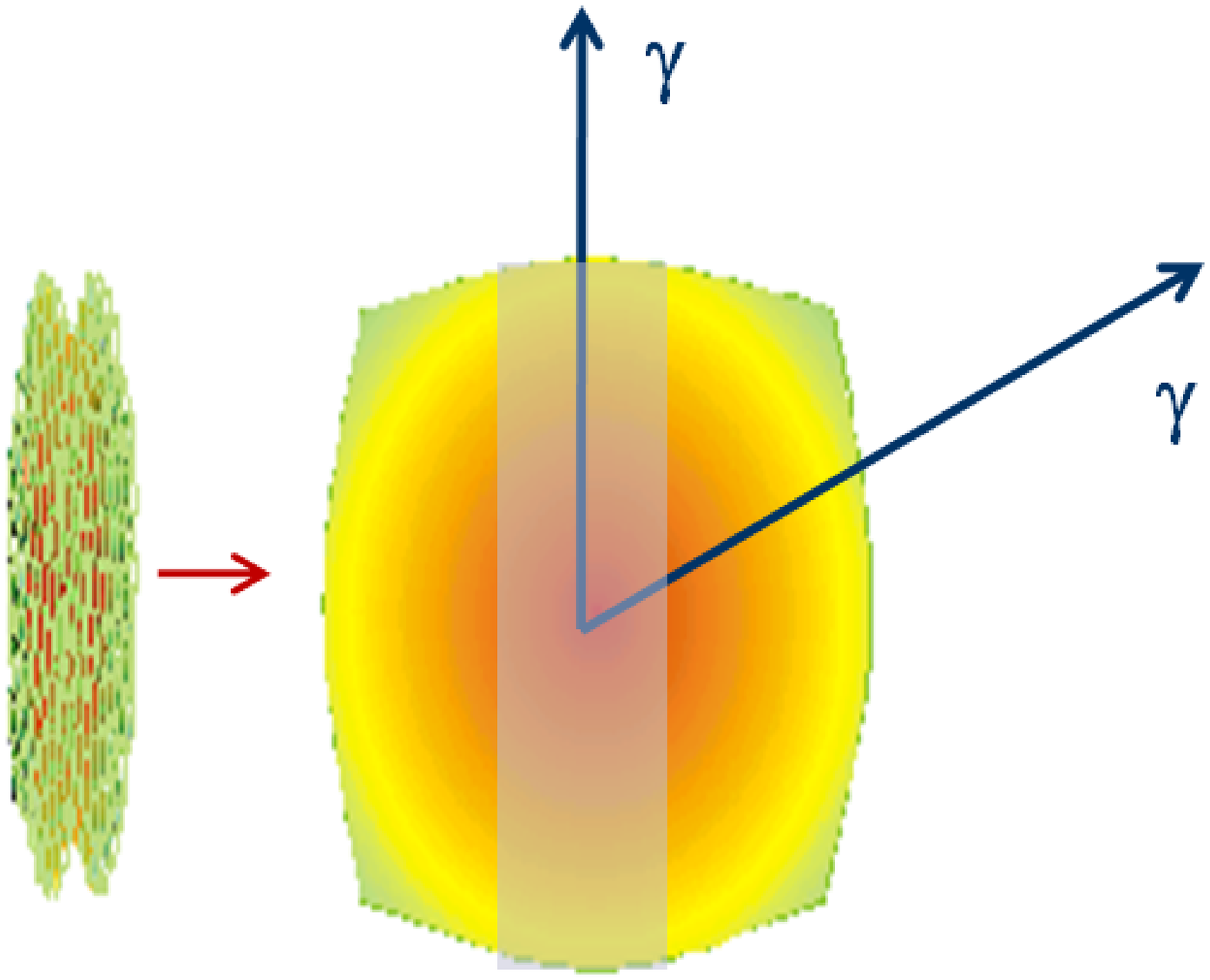}
\caption{\label{fig11}Rapidity dependence of direct photons.}
\end{minipage}
\end{figure}
It is predicted that the second order of the Fourier transfer coefficient
($v_2$, elliptic flow) of angular distributions of photons show the different
sign and/or magnitude, depending on the production processes~\cite{ref21}
(Fig.~\ref{fig10}). The observable is powerful to disentangle the
contributions from various photon sources in the $p_T$ region where they
intermix. The photons from hadron-gas interaction and thermal radiation may
follow the collective expansion of a system, and give a positive $v_2$.
The amount of photons produced by jet-photon conversion or in-medium
bremsstrahlung increases as the medium to traverse increases.
Therefore these photons show a negative $v_2$. The fragmentation photons
will give positive $v_{2}$ since larger energy
loss of jets is expected orthogonal to the event plane.

\begin{figure}[h]
\begin{center}
\includegraphics[width=35pc]{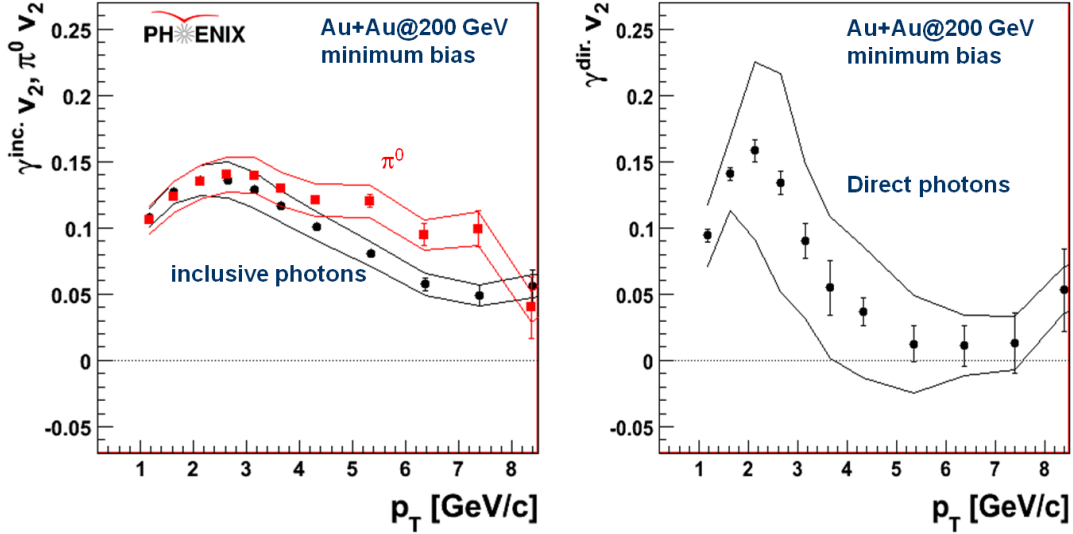}
\caption{\label{fig12}Elliptic flow of (a, left) $\pi^0$ and inclusive photons and (b, right) direct photons.}
\end{center}
\end{figure}

PHENIX has measured the $v_2$ of direct photons by subtracting the $v_2$
of hadron decay photons off from that of the inclusive photons, following
the formula below:
\[ {v_2}^{dir.} = (R \times {v_2}^{incl.} -{v_2}^{bkgd.})/(R-1),\ \ \ R = (\gamma/\pi^0)_{meas}/(\gamma/\pi^0)_{bkgd}\]
The elliptic flow of $\pi^0$ and inclusive photons are shown in
Fig.~\ref{fig12}(a), and the one for direct photons is shown in
Fig.~\ref{fig12}(b).
The $v_2$ of direct photons is large and positive, and comparable to the flow
of hadrons for $p_T<$3\,GeV/$c$. This result is hard to be explained by many
models. Several models qualitatively predicted the positive flow of the
photons assuming the photons are boosted with hydrodynamic expansion of the
system, but the amount is significantly lower than the
measurement~\cite{Chatterjee:2009}. There is one model that gives relatively
large flow by including hadron-gas interaction~\cite{vanHees:2011vb}.

\section{Summary}
Direct photons are a powerful tool to investigate the collision dynamics.
PHENIX has measured direct photons over wide $p_T$ ranges, including hard
scattering and thermal photons, and extracted quantities, such as slope
parameters, that reflect thermodynamic properties of the matter.
An unexpectedly large positive elliptic measured for direct photons are hard
to be explained by many models.

\end{document}